\documentclass{article}
\usepackage{ijcai23}

\DeclareFontFamily{U}{mathx}{\hyphenchar\font45}
\DeclareFontShape{U}{mathx}{m}{n}{
      <5> <6> <7> <8> <9> <10>
      <10.95> <12> <14.4> <17.28> <20.74> <24.88>
      mathx10
      }{}
\DeclareSymbolFont{mathx}{U}{mathx}{m}{n}
\DeclareMathSymbol{\bigtimes}{1}{mathx}{"91}

\usepackage{times}

\usepackage{soul}
\usepackage{amsfonts}
\usepackage{url}
\usepackage[hidelinks]{hyperref}
\usepackage[utf8]{inputenc}
\usepackage[small]{caption}
\usepackage{graphicx}
\usepackage{amsmath}
\usepackage{amsthm}
\usepackage{booktabs}
\usepackage{algorithm}
\usepackage{algorithmic}
\usepackage[switch]{lineno}

\linenumbers

\newtheorem{theorem}{Theorem}

\newcommand*{\Q}{\mathbb{Q}}
\newcommand*{\Z}{\mathbb{Z}}
\renewcommand*{\L}{\mathcal{L}}

\newcommand*{\A}{\mathcal{A}}

\newcommand*{\R}{\mathsf{R}}

\newcommand{\Win}{\mathsf{Win}}

\usepackage{multirow,array}

\usepackage{soul}
\usepackage{url}
\usepackage[hidelinks]{hyperref}
\usepackage{amsmath}
\usepackage{amsthm}
\usepackage{booktabs}
\usepackage{algorithm}
\usepackage{algorithmic}
\usepackage{tikz-cd}
\usepackage{tikz}
\usetikzlibrary{automata, positioning, arrows,shapes, snakes}
\tikzset{
->, 
node distance=2cm, 
every state/.style={thick, fill=gray!10}, 
initial text=$ $, 
}
\usepackage{forest}
\title{Multi-Agent Systems with Quantitative Satisficing Goals}

\author{
Senthil Rajasekaran$^1$\and
Suguman Bansal$^2$\And
Moshe Y. Vardi$^{1}$\\
\affiliations
$^1$Rice University\\
$^2$Georgia Institute of Technology\\
\emails
sr79@rice.edu,
suguman@gatech.edu,
vardi@rice.edu
}
\begin{document}
\nolinenumbers 

\maketitle
\begin{abstract}
In the study of reactive systems, \emph{qualitative} properties are usually easier to model and analyze than \emph{quantitative} properties. This is especially true in systems where mutually beneficial cooperation between agents is possible, such as multi-agent systems. The large number of possible payoffs available to agents in reactive systems with quantitative properties means that there are many scenarios in which agents deviate from mutually beneficial outcomes in order to gain negligible payoff improvements. This behavior often leads to less desirable outcomes for all agents involved. For this reason we study \emph{satisficing} goals, derived from a decision-making approach aimed at meeting a good-enough outcome instead of pure optimization. By considering satisficing goals, we are able to employ efficient automata-based algorithms to find pure-strategy Nash equilibria. We then show that these algorithms extend to scenarios in which agents have multiple thresholds, providing an approximation of optimization while still retaining the possibility of mutually beneficial cooperation and efficient automata-based algorithms. Finally, we demonstrate a one-way correspondence between the existence of $\epsilon$-equilibria and the existence of equilibria in games where agents have multiple thresholds.
\end{abstract}

\section{Introduction}



Work in formal verification has seen recent trends towards generalizations, such as considering \emph{quantitative} properties as opposed to qualitative ones \cite{wATL,guptathesis,kwiatkowska2007quantitative} and reasoning about \emph{multi-agent} systems as opposed to two-agent systems \cite{GTW02,guptathesis,mogavero2014reasoning,SLmultiagentbook,van2008multi,wooldridge2009introduction}. This motivates combining both generalizations to reason about extremely general systems.

The quantitative properties in the multi-agent formal verification literature are usually derived from modifications of qualitative properties, i.e. temporal logics \cite{wATL,slf}. Recent work in fields like reinforcement learning and planning \cite{rlbook1,planningbook1}, however, trends towards a different type of system in which more general quantitative state-based rewards are considered. In these systems, agents choose actions in order to receive rewards, and the quantitative property being considered reasons directly about the sequence of rewards obtained. Since the sum of these rewards is likely to be infinite in an infinite execution of a system, an aggregation function is usually applied to this sequence \cite{Osborne1994,rlbook1}. One of the most common aggregation functions is the \emph{discounted sum} -- as each time step passes all rewards are decreased geometrically by multiplication through a discount factor between $0$ and $1$ \cite{Osborne1994}. This function is popular since it not only guarantees convergence but also discounts rewards in the future against rewards in the present, incentivizing agents to follow the rational economic behavior of preferring to receive rewards immediately. The use of the discounted sum then allows for us to reason about reward sequences, commonly referred to as \emph{payoffs}, in such systems.

We choose to reason about the \emph{Nash equilibrium} solution concept as our quantitative property of interest as it is one of the most important concepts in the theory of multi-agent systems~\cite{Nash48,SLmultiagentbook,wooldridge2009introduction}. A Nash equilibrium consists of a strategy for each agent such that no agent in the system can increase his payoff through a unilateral deviation. Since agents are not incentivized to change their strategy, Nash equilibria represent stable points in multi-agent systems. Deterministic behavior is preferred in many settings (i.e. formal verification), so \emph{pure} Nash equilibria, which consist of deterministic strategies, are a preferable concept~\cite{bouyer2010nash,bouyer2011nash,ummels2015pure}. We adopt this preference here, referring to pure Nash equilibria whenever we write Nash equilibria.

In these systems it is natural to consider agents that are incentivized by maximizing their payoff -- i.e. their \emph{goal} is payoff maximization. A difficulty arises, however, when this type of agent is considered alongside the Nash equilibria solution concept. Agents with a maximization goal are incentivized to deviate for \emph{an arbitrary} improvement in payoff, even if that improvement is minuscule, a situation that naturally arises with the accumulation of a discount factor. We demonstrate that this behavior removes opportunities for mutually beneficial cooperation between agents with an example (\autoref{fig:motivatingexample}). For this reason the notion of an $\epsilon$-equilibria \cite{complexityofnash,roughgarden2010algorithmic} is often considered, in which an agent is not incentivized to deviate unless it gain at least $\epsilon$ more payoff for some fixed $\epsilon$. While this solution concept disincentivizes arbitrary deviations, we are not aware of general solutions to finding $\epsilon$-equilibria in these systems. Thus we are left with a desire for a system that disincentivizes minuscule deviations (increasing cooperative opportunities) and allows for efficient algorithms that find Nash equilibria. 

We address this gap by introducing a new type of agent goal based on the concept of \emph{satisficing} ~\cite{simon1956rational}. An agent with a satisficing goal is only interested in searching for some payoff that would meet a fixed threshold that represents a ``good enough" outcome. The use of this type of goal has been successful in stochastic systems with two agents~\cite{BCVTACAS21,bansal2022synthesis}. Satisficing goals accomplish two things. First, they address the problem of agents deviating for negligible gain. Since agents only have a threshold to meet, deviations are not profitable unless they meet the previously unmet threshold. Second, they transform our quantitative system into one with qualitative properties, allowing us to use efficient automata-based techniques \cite{RV21,RV22}. 

Satisficing goals, however, come at the cost of removing much of the expressiveness we wished to originally capture by considering general quantitative systems. To this end, we introduce a novel approach based on \emph{multi-satisficing} goals. An agent with a multi-satisficing goal has multiple thresholds to meet and is incentivized to meet as many of his thresholds as possible. This allows us to recapture much of the expressiveness of our original quantitative system while still not incentivizing minuscule deviations. Furthermore,  we show how to extend the constructions used for satisficing goals (with single thresholds) to retain the complexity-theoretic benefit of our automata-theoretic approach, maintaining a PSPACE upper bound for finding Nash equilibria that matches the complexity of the qualitative setting~\cite{RV21,RV22}. 
We conjecture that this bound is tight because the problem of finding Nash equilibria requires keeping track of the states of an unbounded number of agents. Since, however, satisficing goals in multi-agent games are a novel concept we found it challenging to construct a complexity-theoretic reduction; the specification of satisficing goals is quite different from the specifications of the canonical PSPACE-hard multi-agent problems~\cite{kozen77,sipser2006,Hopcroftintroduction}. 

Finally, by considering many thresholds that are close together we can establish a relationship between Nash equilibria in systems where agents have multi-satisficing goals to the more traditional notion of $\epsilon$-equilibria. Section 6 demonstrates how to construct thresholds in a system  with multi-satisficing goals such that the existence of an equilibria in that system corresponds to an $\epsilon$-equilibria in the system where only payoffs are considered. The inverse relationship represents a powerful open question, as a positive result would allow for a highly efficient automata-based algorithm that solves for  $\epsilon$-equilibria in a very general form of system.

The main contribution of this this paper is the introduction of a new type of goal (multi-satisficing) for agents in a very general quantitative multi-agent concurrent game setting. Three main reasons are provided in support of these goals. First, they provide a framework that offers more opportunities for collaboration between agents, providing a strategic reason to consider multi-satisficing goals. Second, they admit efficient automata-based algorithms. In Section 4, we discuss how a naive application of the techniques in \cite{BBMU15} would yield a NEXPTIME upper bound as opposed to the PSPACE upper bound provided in our paper, providing a complexity-theoretic reason to consider this approach to multi-satisficing goals. Finally, this paper demonstrates a one-way correspondence between equilbiria in systems with multi-satisficing goals and $\epsilon$-equilibria, tying the concept of multi-satisficing goals back to a widely accepted and used notion in the literature.

\section{Preliminaries}

\paragraph{B\"uchi automata, Co-safety automata and Safety automata.}
A  {\em B\"uchi automaton} is a tuple
  $\A = \langle  \Sigma , Q, q_0, \delta, F \rangle$, where  $ \Sigma $ is a finite {\em input alphabet}, $ Q $ is a finite set of {\em states}, $ q_0 \in Q $ is the {\em initial state}, $ \delta \subseteq (Q \times \Sigma \times Q) $ is the   {\em transition relation},  and $ F \subseteq Q $ is the set of {\em accepting states}.
A B\"uchi automaton is {\em deterministic} if for all states $ q $ and input symbols $s \in \Sigma$, we have that $ |\{q'~:~(q, s, q') \in \delta  \}| \leq 1 $.
For a word $ w = w_0w_1\dots \in \Sigma^{\omega} $, a {\em run} $ \rho$ of $\A$ over $ w $ is a sequence $q_0,q_1\dots\in Q^\omega$ such that $ q_0$ is the initial state  and $ \tau_i =(q_i, w_i, q_{i+1}) \in \delta $ for all $i \in \mathbb{N}$.
Let $ \mathit{inf}(\rho) $ denote the set of states that occur infinitely often in run ${\rho}$. 
A run $\rho$ is an {\em accepting run} if $ \mathit{inf}(\rho)\cap F \neq \emptyset $. A word $w$ is  accepted by $\A$ if $\A$ has an accepting run over $w$. The \emph{language} of $\A$, denoted $\L(\A)$ is the set of words accepted by $\A$. Languages accepted by B\"uchi automata are called $\omega$-regular and are closed under set-theoretic union, intersection, and complementation~\cite{GTW02}. A {\em safety automaton} is a deterministic B\"uchi automaton with a single rejecting sink state~\cite{kupferman1999model}. Such an automaton accepts a word $w$ if no finite prefix of $w$ leads to the rejecting state.
A {\em co-safety automaton} is a deterministic B\"uchi automaton with a single accepting sink state~\cite{kupferman1999model}. Such an automaton accepts a word $w$ if some finite prefix of $w$ leads to the accepting state.

\paragraph{Comparator automata.}
Given an aggregate function $f:\Z^{\omega}\rightarrow \mathbb{R}$, a relation $\mathsf{R} \in \{<, >, \leq, \geq, =, \neq\}$, and a threshold value $v \in \Q$, the {\em comparator automaton for $f$ with upper bound $\mu\in \Z$, relation $\mathsf{R}$, and threshold $v\in\Q$} is an automaton that accepts an infinite word $A$ over the alphabet $\Sigma=\{-\mu,-\mu+1,\ldots \mu\}$ iff $f(A)$ $ \mathsf{R}$ $v$ holds~\cite{BCVFoSSaCS18}. The discounted sum of an infinite-length weight-sequence $W = w_0,w_1,\dots \in \Z^\omega$ with discount factor $\gamma>1$ is given by $\sum_{i=0}^\infty \frac{w_i}{\gamma^i}$. The comparator automaton for the discounted-sum has been shown to be a safety or co-safety automaton when the discount factor $\gamma>1$ is an integer for all values of  $\R$,  $\mu$ and $v$~\cite{BVCAV19,BCVTACAS21}. It is further known that there are no B\"uchi comparator automata for non-integer  discount factors $\gamma>1$~\cite{vardi2022comparator}. 
Here, we only consider integral discount factors.

\paragraph{Multi-agent Systems.}
Multi-agent systems provide a powerful framework for reasoning about interactions between multiple strategic agents. One of the most general multi-agent systems is given by a \emph{concurrent game}, defined as a 6-tuple $\langle V, {\bf v_0} ,\Omega, A, \tau, \alpha \rangle$, where ({\bf 1}) $V$ is a finite set of states. ${\bf v_0} \in V$ is the initial state ({\bf 2}) $\Omega = \{0 \ldots k-1\}$ is a set of agents of size $k$ ({\bf 3}) $A = \langle A_0 \ldots A_{k-1}\rangle $ is a tuple of action sets; the set $A_i$ is associated with Agent $i$, and it represents the actions available to that agent. We denote $  D = A_0 \times A_1 \times \ldots \times A_{k-1}$ as the set of \emph{decisions}, which represent global actions by the full set of agents ({\bf 4}) $\tau : V \times D \rightarrow V$  is the \emph{transition function}.  ({\bf 5}) $\alpha = \langle\alpha_0 \ldots \alpha_{k-1}\rangle $ is a tuple of \emph{goals}. The goal $\alpha_i$ is associated with Agent $i$, and it represents some ordering on outcomes of the game that constitutes the agent's preferences.  We discuss different types of goals below.
Intuitively, at each state $v \in V$ of the game, the agents concurrently choose actions from their set of actions.  These actions then determine a decision, which is used to transition the state of the game via $\tau$. 
Given this framework, we now define the notion of a \emph{play} in a concurrent game. A play $\rho \in (V \times D)^{\omega}$ is an infinite sequence of states and decisions  $({\bf v_0}, d_0), (v_1, d_1) \ldots$ such that ${\bf v_0}$ is the initial state and, for all $j\geq 0$, $ v_{j+1} = \tau(v_j,d_j)$. Plays represent outcomes of the game, so each $\alpha_i$ represents an ordering on the set of plays as we discuss further below.

Agents are then incentivized to choose actions that create plays with maximal preference. These actions are dictated by a \emph{strategy}. A strategy for Agent $i$ is a function $\pi_i : (V\times D)^* \times V \rightarrow A_i$. Intuitively, this is a  deterministic function that, given the observed history of the game (represented by a sequence of previously seen state-decision pairs $(V\times D)^*$  and a current state $V$) returns an action $a_i \in A_i$.


Let $\Pi_i$ represent the set of strategies for agent $i$. The set of \emph{strategy profiles} is defined as 
$ \Pi = \bigtimes_{i \in \Omega} \Pi_i$. Thus, a strategy profile $\pi \in \Pi$ is a tuple of strategies, one for each agent, of type $(V\times D)^* \times V \rightarrow D$. We define $\pi_{-i}$ to be the strategy profile $\pi$ with the $i$-th element removed. By a slight abuse of notation, we combine $\pi_{-i}$ with a strategy $\pi'_i$ for Agent $i$ to create a new strategy profile $\langle \pi_{-i},\pi'_i \rangle$. 

Strategies are deterministic, so given a concurrent game $G=\langle V, {\bf v_0} ,\Omega, A, \tau, \alpha \rangle$ and a strategy profile $\pi$, there is a unique play $\rho$ resulting from a strategy profile $\pi$, given by
({\bf 1}) $d_0=\pi((\epsilon,{\bf v_0}))$ ({\bf 2}) $\rho_0 = \langle {\bf v_0}, d_0 \rangle$ ({\bf 3}) $v_{i+1}=\tau(v_i,d_i)$ ({\bf 4}) $d_{i+1}=\pi((\rho_0,\ldots,\rho_i),v_{i+1})$ ({\bf 5}) $\rho_{i+1}=\langle v_{i+1},d_{i+1}\rangle$. We denote this play as $\rho_{\pi}$ and call it the \emph{primary trace} of $\pi$.

We now define the notion of a \emph{Nash equilibrium}. A Nash equilibrium is a strategy profile $\pi = \langle \pi_0 \ldots \pi_{k-1} \rangle $ such that for every agent $i \in \Omega$ there is no Agent~$i$ strategy $\pi'_i$ such that, under the preference induced by $\alpha_i$, the play $\rho_{ \langle \pi_{-i}, \pi'_i \rangle}$ is strictly preferred to the play $\rho_{\pi}$. Therefore, under a Nash equilibrium strategy profile no agent may deviate from the profile in order to obtain a more preferable result. Since both the notion of strategies for individual agents and the transition function in a concurrent games are deterministic here, when we refer to concepts such as Nash equilibria, we implicitly mean \emph{pure-strategy} Nash equilibria.

\paragraph{Turn-Based games.}
An important class of games are \emph{turn-based} games, in  which the state set $V$ is partitioned among the $k$ agents: $V=V_0 \cup \ldots \cup V_{k-1}$. In a state $v\in V_i$ only the action of Agent~$i$ affects the transition. Formally, let $d$ and $d'$ be two decisions such that $d[i]=d'[i]$ (they agree on $A_i$), then we have that $\tau(v,d)=\tau(v,d')$. It is convenient to view this as if only Agent~$i$ takes an action $a$ in a state $v\in V_i$, and, by slight abuse of notation we write that this action causes a move from the state $v$ to the state $\tau(v,a)$. We denote turn-based games as $\langle V_0,\ldots,V_{k-1}, {\bf v_0},E, \alpha \rangle$, where $E$ is a set of edges over $V = \bigcup_{i=0}^{k-1}V_i$ describing the possible moves in the game, the sets of agents and actions are implicit.

\paragraph{Reachability and Safety games.}
Two important types of goals $\alpha_i$ are specified by \emph{Reachability} and \emph{Safety} conditions. A reachability goal is specified by a set of states $ R \subseteq V$, and an agent with a reachability goal prefers plays that visit a state $r \in R$ over those that do not. A safety goal, which can be seen as the dual of a reachability goal, is specified by a set of states $S \subseteq V$, and an agent with a safety goal prefers plays that do not visit states in $V \setminus S$ over those that do. 

In this paper, when we refer to reachability and safety \emph{games} we specifically mean a two-agent turn-based game in which one agent has a reachability or safety goal and the other agent has the negation of this goal \cite{McNaughton1993InfiniteGP}.  In a reachability game $G = \langle V_0,V_1, {\bf v_0},E,R \rangle$ Agent~$0$ has a preference for plays that visit at least one state in the reachability set specified by $\alpha_0=R$; Agent~$1$ has the safety goal $\alpha_1=V\setminus R$ so the two goals are mutually exclusive.  

For both reachability and safety games we use the notation $G = \langle V_0 , V_1 , E, C\rangle $, where $C$ represents the winning condition (either a reachability or safety set) without specifying an initial state. Instead, we partition the states $V$ into two sets. We say a state $v \in V$ is in $\Win_0(G)$ if Agent~$0$ has a \emph{winning strategy} $\pi_0$ starting in state $v$, which means that for all Agent~$1$ strategies $\pi_1$, we have that $\rho_{\langle \pi_0, \pi_1 \rangle}$ is winning for Agent~$0$ when $v$ is treated as the initial state. If no such Agent~$0$ strategy exists then  Agent~$1$ has a winning strategy from $v$ and $v$ belongs to the analogously defined $\Win_1(G)$. Both sets can be computed in time $O(|V|+|E|)$~\cite{Boreldeterminancy,McNaughton1993InfiniteGP}.

\section{Satisficing  Games}

A \emph{discounted-sum game}  $\langle V, {\bf v_0} ,\Omega, A, \tau, \gamma, \mathcal{W}, \alpha \rangle$ is a concurrent game $\langle V, {\bf v_0} ,\Omega, A, \tau, \alpha \rangle$ with the following additions:
({\bf 1}) $\gamma > 1$ is an integral discount factor, and 
({\bf 2}) $\mathcal{W} : V \times D \rightarrow \mathbb{Z}^k$ is a \emph{reward function} that associates each state-decision pair with a $k$-length integer-valued vector. 
Intuitively, the reward function associates integer rewards for each agent as a function of the state and the decision. The discount factor devalues rewards in the future with respect to rewards in the present, disincentivizing agents from waiting before taking actions to get rewards. The addition of these quantitative properties allows us to assign numerical values to plays, which we use to define goals $\alpha$. Given a play $\rho = ({\bf v_0}, d_0), (v_1, d_1) \ldots$, the \emph{cumulative reward} earned by Agent $i$ from $\rho$ is defined as $R_i(\rho) = \sum_{j=0}^{\infty} \mathcal{W}(v_j,d_j)[i] \cdot \frac{1}{\gamma^{j}}$.

An often-used choice for the goal $\alpha_i$ is an \emph{optimization} goal, i.e. a maximization goal which always prefers plays with higher cumulative reward to those with lower cumulative reward.  An alternative we consider here is a \emph {satisficing goal} $\alpha_i$, specified by a pair $\langle \mathsf{R}_i, t_i \rangle$, where  $\mathsf{R}_i \in \{ <, \leq, > , \geq, = , \neq\}$ is a comparison relation and $t_i \in \mathbb{Q}$ is a threshold. These goals induce a binary preference on plays - an agent with a satisficing goal prefers plays $\rho$ in which $R_i(\rho)~\mathsf{R}_i~t_i$ holds to plays where it does not hold. We demonstrate the utility of considering agents with satisficing goals as opposed to optimization goals under the Nash equilibria solution concept through an example.
\begin{figure}[t]
\centering
\begin{tikzpicture}
\node[state, circle, initial] (q1) {$q_1$};
\node[state, diamond, below left of=q1] (q2) {$q_2$};
\node[state, diamond, below right of=q1] (q3) {$q_3$};
\node[state, diamond, right of=q3] (q4) {$q_4$};

\draw (q1) edge[above left]node{$[n,0]$} (q2)
(q1) edge [above right] node{$[0,0]$} (q3)
(q2) edge [loop below] node{$[0,0]$} (q2)
(q3) edge [loop below] node{$[0,1]$} (q3)
(q3) edge [below] node{$[m,0]$} (q4)
(q4) edge [loop below] node{$[0,0]$} (q4)

;
\end{tikzpicture}
\caption{A two-agent turn-based discounted-sum game with states $V = \{q_1,q_2,q_3,q_4\}$. Agent~$0$ owns the circled state ($V_0 = \{q_1\}$); Agent~$1$ owns the diamond states ($V_1 = \{q_2,q_3,q_4\}$).}
\label{fig:motivatingexample}
\end{figure}
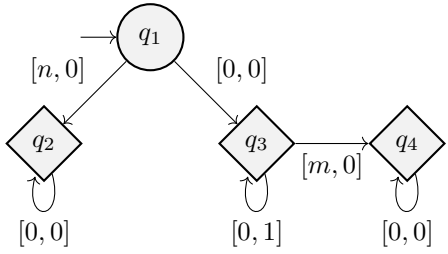

Consider the two-player turn-based discounted-sum game in Fig~\ref{fig:motivatingexample}. Each edge is labeled with a pair of integers, representing the reward function $\mathcal{W}$ - Agent~$0$ and Agent~$1$ receive the first and second reward, respectively.  Suppose that $ m >> n$, such that for our given discount factor $\gamma$ we have $\frac{m}{\gamma^{k}} > n > \frac{m}{\gamma^{k+1}} $ for a large $k$. This means there exists a path from $q_1$ to $q_4$ that loops in $q_3$ for several turns such that the (discounted) reward gained by Agent~$0$ is higher than what it would have achieved on the path from $q_1$ to $q_2$.


Assume that both agents are given maximization goals. Note that  the single play that goes from $q_1$ to $q_2$ and stays there represents the only Nash equilibrium in this game. In this strategy, Agent~$0$ chooses to go to $q_2$ leading to a reward of $n$ and $0$ for both  agents, respectively. The alternative Agent~$0$ strategy of moving to $q_3$ would lead to a 0 reward for the Agent~$0$, since Agent~$1$ is incentivized to loop at $q_3$ forever -- even when the marginal reward gained by each successive loop is minuscule due to the discount factor. This is because Agent~$1$ always chooses plays with more cumulative reward given according to her maximization goal $\alpha_i$. 
Thus, the game's only stable point emerges from a double threat - Agent~$1$ threatens to loop at $q_3$ forever, while Agent $0$ does not give her the chance, as Agent~$1$'s desire to loop at $q_3$ forever for increasingly minuscule rewards means that Agent~$0$ cannot ``negotiate'' with her.


For this reason, a notion of $\epsilon$-Nash equilibria is sometimes considered, in which a deviation is not considered preferable unless it gains at least $\epsilon$ more reward for some fixed constant $\epsilon > 0$ \cite{complexityofnash,roughgarden2010algorithmic}. This gives the players a sense of {\em rationality} in that they will not deviate for arbitrarily minuscule rewards. Here, we choose another form of rationality based on attaining certain thresholds via satisficing goals. Satisficing is a decision-making strategy that entails searching through available alternatives until a {\em good-enough} threshold is met \cite{simon1956rational}. It is, therefore, a reasonable approach by a decision-maker in circumstances where an optimal solution cannot be determined or are undesirable. In our example, if Agent~$1$ had a properly chosen satisficing goal instead of an optimization goal, she could accept a strategy that loops at $q_3$ a finite number of times before moving to $q_4$. To this end we study the existence of Nash equilibria in \emph{satisficing games}, which are discounted-sum games where each agent has a satisficing goal. 

We note that the change from optimization to satisficing goals also influences the definition of Nash equilibria since there are only two possible preferences for each agent. Given a fixed play $\rho$ in a satisficing game, for each Agent $i$ it is either the case that $R_i(\rho)~\mathsf{R}_i~t_i$ holds or does not. Thus, we introduce the concept of a \emph{winning set} $W$ induced by a play $\rho$ to be the set of agents $W \subseteq \Omega$ such that $i \in W$ iff $R_i(\rho) \alpha_i$ holds. 
Therefore we can find Nash equilibria by finding \emph{W-Nash Equilibria} ($W$-NE for short), which are Nash equilibria with winning set $W$ - varying $W$ as needed. 

The next section develops the technical tools needed to find $W$-NE in satisficing games. Satisficing games are able to avoid certain undesirable behaviors from agents, but this comes at the cost of equipping agents with only two possible payoffs. While the tendency for agents with optimization goals to deviate for negligible rewards stemmed from the potentially uncountable number of payoffs available, limiting agents to two possible payoffs in the satisficing case presents its own concerns. In Section 5 (Multi-Satisficing Goals) we show how to extend the analysis of satisficing goals to allow for a finite but unbounded number of payoffs for each agent, which we believe represents an important middle ground between satisficing and optimization goals.


\section{Characterization of Nash Equilibria}
Recall that agents in multi-agent satisficing games distinguish between two types of plays -- those that satisfy their goal and those that do not. Therefore, we can consider the existence of Nash equilibria by considering $W$-Nash equilibria. Our goal is to develop an automata-theoretic characterization of $W$-NE existence; we begin by describing how to model satisficing goals via automata.

Since the discount factor $\gamma$ is integral, we construct  comparator automata that recognize infinite sequences of rewards that satisfy a given satisficing goal in a multi-agent satisficing game. Let $\mu_i$ represent the maximum magnitude of the rewards assigned to Agent~$i$, i.e. the range of $\mathcal{W}(\cdot)[i]$ is in the interval $[ -\mu_i,\mu_i ]$. By following the construction in \cite{BCVTACAS21}, we construct a comparator automaton $A^i$ that recognizes reward sequences that satisfy the goal $\alpha_i = \langle \mathsf{R}_i, t_i \rangle$. The size of $A^i$ is $\frac{\mu_i \cdot \eta_i}{\gamma-1}$, where $\eta_i$ is the length of the base $\gamma$ representation of $t_i = t_i[0] \ldots (t_i[m] \ldots t_i[\eta-1])^{\omega}$. This is a linear-sized construction under the assumption that $\mu_i$ is provided in unary. Furthermore, if $\mathsf{R}_i \in \{ \leq , \geq, = \}$ then $A^i$ is a safety automaton; it is a co-safety automaton otherwise~\cite{BCVTACAS21}. Therefore when given a  satisficing goal $\alpha_i$, we construct a B\"uchi automaton $A^i = \langle  \Sigma^i , Q^i, q^i_0, \delta^i, F^i \rangle$, where $\Sigma^i = \{ -\mu_i \ldots \mu_i \}$. Note that when $A^i$ is a co-safety automaton, once a state in $F^i$ is reached the automaton accepts and when $A^i$ is a safety automaton, once a state not in $F^i$ is reached the automaton rejects. Since agents may either have a safety or co-safety automaton representation of their goal, we denote $\Omega_S$ as the set of agents with a safety goal representations; $\Omega_C$ is analogously denoted for co-safety representations. Furthermore, $W_S = \Omega_S \cap W$,$W_C = \Omega_C \cap W$, $\overline{W} = (\Omega\setminus W)$, $\overline{W}_C =  \overline{W} \cap \Omega_C$, and $\overline{W}_S =  \overline{W} \cap \Omega_S$.

The qualitative nature of satisficing goals means that a strategy profile $\pi$ is a $W$-NE iff ({\bf 1}) Precisely the goals for agents in $W$ are satisfied on $\rho_{\pi}$ and ({\bf 2}) For agents $j \not \in W$, there is no Agent-j-strategy $\pi'_j$ such that Agent~$j$'s goal is satisfied on $\rho_{\langle \pi_{-j}, \pi'_j \rangle}$.This is because in the qualitative setting, agents in $W$ have achieved their maximal preference and therefore have no incentive to deviate. Therefore, it is enough to only consider deviations from agents outside of $W$.  Following~\cite{RV21}, we refer to the first condition as the Primary-Trace condition and the second as the $j$-Deviant-Trace Condition. We proceed by analyzing each condition separately and then show how to unify them.  

\subsection{The Primary-Trace Condition}

In order to analyze the Primary-Trace Condition for a given set $W$, we construct a B\"uchi automaton $A_W = \langle D, Q, q^W_0, \delta_W, F_W \rangle$ that recognizes words $ \rho \in D^\omega$ iff they correspond to plays that satisfy the Primary-Trace Condition. Note that since the transition function in a multi-agent satisficing game is deterministic and the initial state ${\bf v_0} \in V$ is fixed, plays are uniquely determined by an infinite sequence of decisions, so we take the input alphabet to be $D$.  The other components of $A_W$ are defined as follows: ({\bf 1}) $Q = V \times \bigtimes_{i\in \Omega} Q^i \times 2^{\Omega} \times 2^{\Omega} $ ({\bf 2}) $q^W_0 = \langle{\bf v_0} , q^0_0 \ldots q^{k-1}_0, W_C , \overline{W}_S\rangle$ ({\bf 3}) For $q = \langle v, q_0 \ldots q_{k-1}, S_1, S_2 \rangle$, with $S_1,S_2\subseteq\Omega$  and $d$ a decision, the transition $\delta^W(q,d)$ is  $\langle v', q_0' \ldots q_{k-1}', S_1',S_2'\rangle$, where ({\bf 3.i}) $v' = \tau(v,d)$ ({\bf 3.ii}) For $i \in \Omega$, we have $q_i' = \delta^i(q_i, \mathcal{W}(v,d)[i])$ ({\bf 3.iv}) Finally, we specify how we change $S_1$ and $S_2$ to $S'_1$ and $S'_2$ by removing states -- we always have $S'_i \subseteq S_i$: ({\bf 3.iv.a}) If $i \in S_1$ then $i \not \in S_1'$ if $q_i' \in F^i$. ({\bf 3.iv.b}) If $i \in S_2$, then $i \not \in S_2'$ if $q_i' \not \in F^i$. ({\bf 3.iv.c}) If $i \in W_S$, we specify that the transition is not defined if $q_i' \not \in F^i$. Upon attempting such a transition the automaton $A_W$ rejects. ({\bf 3.iv.d})If $i \in \overline{W}_C$, we specify that the transition is not defined if $q_i' \in F^i$. Upon attempting such a transition the automaton $A_W$ rejects. ({\bf 4}) $F_W$ is the set of states with $S_1 = S_2 = \emptyset$.
        

Intuitively, $S_1$ represents agents in $W$ with co-safety goals that have yet to be satisfied. Once a final state for such a goal has been reached, this goal is satisfied. A dual logic applies to $S_2$, which represents agents not in $W$ with safety goals. At some point such goals must reach their rejecting state. For co-safety goals for agents not in $W$, we must make sure they never reach  a final state, so we immediately reject if they do. The same holds for safety goals in $W$, if we reach a rejecting state then $A_W$ must also reject.

\begin{theorem}[$A_W$ Correctness]
For a given multi-agent satisficing game $\mathbb{G}$, $A_W$ accepts a word $\rho \in D^{\omega}$ if $R_i(\rho) \textrm{ } \mathsf{R}_i \textrm{ } t_i$ holds for precisely the agents $i \in W$.
\end{theorem}

\begin{proof}

We  break this proof out into four simple cases. First, if $i \in W_C$, then we make sure that $A^i$ reaches a final state before $A_W$ accepts. Due to the definition of $F$ and the initial state $q^W_0$, a final state in $A_W$ cannot be reached unless a final state in $A^i$ is reached somewhere along the way. Due to the cosafety condition on $A^i$ we know that once in a final state $A^i$ will remain in its final states and therefore accept. Second, if $i \in \overline{W}_S $ then we make sure that the safety condition is broken at some point during the run of $A_W$ by the definition of $F$ and the initial state $q^W_0$. $A_W$ cannot enter a final state until this requirement is met. Third, if $i \in W_S$, then we make sure that $A^i$ never leaves its safety set $F^i$. When it does, we specify that such transitions are undefined and halt the automaton. Therefore there can never be an accepted word $d \in D^W$ for $A_W$ that leaves the safety set for $A^i$. Finally, if $i \in (\Omega\setminus W)_C$, then we make sure that $A^i$ never enters a final state on accepting runs of $A_W$. So, these goals cannot be satisfied since their final states are never reached.

\end{proof} 

\subsection{The $j$-Deviant-Trace Condition}

In order to analyze the $j$-Deviant-Trace Condition for an agent $j$ we wish to characterize the set of states $(v, q^j)$ and $(v, q^j, d)$ with $v \in V, q^j \in Q^j, d \in D$ from which there exists some Agent~$j$ strategy that leads to a successful deviation.  To characterize these states we create a two-agent turn-based game $G_j = \langle V_0, V_1, E, C \rangle$ where ({\bf 1}) $V_0 = V \times Q^j$, $V_1 = V \times Q^j \times D$ and ({\bf 2})  For $(v,q^j) \in V_0$, we have $\langle(v, q^j),(v, q^j, d)\rangle \in E$ for all $d \in D$. For $d \in D$, let $d[-j]$ represent $d$ with $A_j$ projected out. We have $\langle(v, q^j,d),(v', q^{j'})\rangle \in E$ iff $\exists d' \in D$ s.t. $d[-j] = d'[-j]$, $\tau(v,d') = v'$, and $\delta^i (q^j,\mathcal{W}(v,d')[j]) = q^{j'}$

The condition $C$ depends on whether Agent~$j$ has a cosafety or safety goal. If Agent~$j$ has a cosafety goal, $C$ is a safety condition specified by the set $V \times Q^j\setminus F^j \cup V\times Q^j\setminus F^j \times D$. If Agent~$j$ has a safety goal, then $C$ is a reachability condition specified in the exact same manner. Intuitively, Agent~$0$ takes control of the agents who are not $j$ and tries to ensure that no successful deviation from $j$ is possible. Therefore, when Agent~$j$ has a cosafety goal Agent~$0$ tries to prevent Agent~$1$ (the stand-in for Agent~$j$) from reaching an accepting state. And when Agent~$j$ has a safety goal Agent~$0$ tries to reach the non-accepting state in the corresponding safety automaton $A^j$. 

\begin{theorem}
Agent $j$ can only successfully deviate from a profile $\pi$ from a state in $\Win_1(G_j)$.
\end{theorem}

That is to say, for a strategy profile $\pi$, if $\rho_{\pi}$ has a prefix $( {\bf v_0},d_0 ) \ldots$ $(v_n, d_n)$ and the run of $A^j$ on the corresponding reward sequence of this prefix puts it in a state $q^j$ such that either $(v_n,q^j)$ or  $(v_n, q^j, d_n)$ belongs to $\Win_1(G_j)$ then Agent~$j$ has a strategy to ensure a successful deviation; otherwise he does not and cannot deviate successfully. 

\begin{proof}
Assume that the game $G_j$ is currently in state $v \in \Win_1(G_j)$. By the definition of $\Win_1(G_j)$, this means that there exists some Agent~$1$ strategy $\pi_1$ that is winning in $G_j$. By following this strategy in $\mathbb{G}$, Agent~$j$ will be able to create a trace that satisfies her goal. If this were not possible and there was some strategy for the other agents $\pi'$ such that $\langle \pi' , \pi_1$ did not satisfy $A^j$ then it would be Agent~$0$ who had the winning strategy $\pi'$ in $G_j$, contradicting the assumption that $v \in \Win_1(G_j)$. Showing that Agent~$j$ cannot deviate successfully deviate from a state in $\Win_0(G_j)$ is entirely similar; now it is $\pi'$ that is the winning strategy in $G_j$.
\end{proof}

\subsection{Characterization of $W$-NE Existence}
With the characterization of  deviations given above, we create the automaton $A'_W =\langle D, Q', q^W_0, \delta'_W, F_W \rangle$, defined as $A_W$ with a restricted state space and transition function. Here, $Q'$ consists of states $\langle v, q_0  \ldots q_{k-1},S_1,S_2 \rangle$ such that $\forall j \not \in W$ we have $\langle v, q_j \rangle \not \in \Win_1(G_j)$. Furthermore, for $q \in Q, d \in D$, $\delta'_W(q,d)$ is now undefined if $\exists j \not \in W$ such that $\langle q[0], q[j], d \rangle \in \Win_1(G_j)$ (Note $q[0] \in V$). Our characterization is then given by the following:
\begin{theorem}
For a given multi-agent satisficing game $\mathbb{G}$, $A'_W$ is nonempty iff there exists a $W$-NE strategy profile.
\end{theorem}
\begin{proof}
($\rightarrow$) Assume that $A'_W$ is nonempty. Then, it accepts at least one word $\rho \in D^{\omega}$. We now create a $W$-NE strategy profile $\pi: (V \times D)^* \times V \rightarrow D$. Recall that since both strategies and the transition function of $\mathbb{G}$ are deterministic and ${\bf v_0}$ is fixed, strategy profiles of type  $\pi: (V \times D)^* \times V \rightarrow D$ have a bijective correspondence to functions of type $D^* \rightarrow D$.
The primary trace of $\pi$ is given by $\rho$ (recalling that sequences of decisions uniquely correspond to plays). Since $A'_W$ accepts on $\rho$, we have that on $\rho$ only the goals in $W$ are satisfied. For an agent $i$ in $W_C$, $A'_W$ can only reach  a final state if a final state in $A^i$ is reached. For an agent $i$ in $W_S$, $A'_W$ gets stuck upon reading a transition to the reject state in $A^i$. The opposite holds for agents $j \not \in W$, so the Primary-Trace Condition holds on $\rho$.

This leaves the $j$-Deviant-Trace Condition. By the construction of $A'_W$, a state that contains a pair $\langle v, q^j \rangle$ for $v \in V$, $q^j \in Q^j$ and reads letter $d \in D$  such that the triple $\langle v, q^j, d \rangle 
\in \Win_1(G_j)$ for some $j \not \in W$ can never be reached, as such transitions are explicitly forbidden. Therefore, should an agent $j \not \in W$ deviate from $\rho$, he must deviate to a state containing a pair $\langle v, q^j \rangle \in \Win_0(G_j)$. This means that from the state corresponding to this pair in $G_j$, Agent $0$ has a winning strategy $\pi_0$ which ensures that for every Agent $1$ strategy $\pi_1$ $A^j$ cannot accept on $\rho_{\langle \pi_0, \pi_1 \rangle}$. From this point on, it is simply a matter of following $\pi_0$ to continue the strategy profile. By following $\rho$ and then the appropriate winning strategy $\pi_0$ for any observed deviations, we are able to create a strategy profile $\pi$ that satisfies the Primary-Trace and $j$-Deviant-Trace Conditions; it is therefore a $W$-NE.

($\leftarrow$)
Now assume that there exists a  $W$-NE strategy profile $\pi$ in $\mathbb{G}$. We show that $A'_W$ accepts $\rho$, the sequence of decisions given along the primary trace of $\pi.$ We claim that $A'_W$ does not get stuck upon reading $\rho$. The first way for $A'_W$ to get stuck is to visit a state containing a pair $\langle v, q^j\rangle $ and read a letter $d$ such that the triple $\langle v, q^j , d \rangle \in \Win_1(G_j)$ for some $j \not \in W$. We note that from these states Agent $1$ has a winning strategy $\pi_1$ in $G_j$. By following this strategy in $\mathbb{G}$, Agent $j$ is able to ensure a play $\rho'$ on which $A^j$ accepts. This corresponds to a violation of the $j$-Deviant-Trace Condition, and so $\rho$ cannot pass through one of these states if $\pi$ is a $W$-NE. The second way for this to happen is if some goal in $W$ is not satisfied on $\rho$ or some goal not in $W$ is satisfied on $\rho$. Either one of these would correspond to a violation of the Primary-Trace Condition, and therefore this contradicts the assumption that $\pi$ was a $W$-NE as well. Furthermore, since the Primary-Trace Condition is satisfied on $\rho$, we have that $A'_W$ reaches a final state and therefore accepts on $\rho$.
\end{proof}

Therefore, we are able to reduce the problem of $W$-NE existence to that of non-emptiness in a B\"uchi automaton. We now analyze the complexity of determining the nonemptiness of $A'_W$. As mentioned before, if the rewards are specified in unary, then each comparator is linear in the size of the input of a multi-agent satisfying game. $A_W$, which consists of the cross-products of these comparator automata and $2^{\Omega}$ is exponential in the size of the input.

In order to determine $\Win_1(G_j)$ for each $j \not \in W$, we either construct a reachability or safety game as outlined in Section 4.2. The size of these games is  $(|V| \times |Q^j| \times |D|) + (|V| \times |Q^j|)$, which is polynomial in the size of the input. These games can be solved in time linear to the size so this step takes polynomial time. Testing the final $A'_W$ therefore lies in NPSPACE=PSPACE, as B\"uchi automata can be tested for nonemptiness in NLOGSPACE \cite{VW94}.

\begin{theorem}\label{W-NE}
The problem of deciding whether a $W$-NE exists in a multi-agent satisficing game is in PSPACE.
\end{theorem}
The characterization of Theorem~\ref{W-NE} is for a fixed $W\subseteq\Omega$. We can also employ the algorithm  to find some $W\subseteq\Omega$, or some $W\subseteq\Omega$ of maximal size such that a $W$-NE exists while remaining in PSPACE. As mentioned in the introduction, one could construct a game consisting of the cross-product of each comparator automaton and the underlying concurrent game in order to apply the algorithms of \cite{ummels2015pure}. This construction is exponential in the size of the input due to the size of the cross-product. Applying the NP algorithm in \cite{ummels2015pure} then yields a NEXPTIME upper bound.


\section{Multi-Satisficing Goals}


In Section 3 we demonstrated a simple scenario in which the tendency for agents with optimization goals to deviate for negligible payoff improvements disallowed the possibility of mutually beneficial cooperation between agents. Satisficing goals were able to address this problem at the cost of allowing each agent only two possible payoffs. Ideally, we would like a goal type that allows agents more than two types of payoffs but still keeps the possibility of mutual beneficial cooperation that is often lost when using optimization goals. In this section we describe a new type of goal in which each agent is equipped with a single relation but has multiple thresholds and payoffs that increase with the number of thresholds satisfied. This allows us to combine the algorithmic efficiency and possibilities for cooperation achieved by using single-threshold satisficing goals with the expressive payoff structure of optimization-type goals.

To this end, we consider \emph{multi-threshold satisficing goals} (henceforth, multi-satisficing goals), which are once again specified by a pair $\langle \mathsf{R}_i, T_i \rangle$.  The difference is that instead of a single threshold $t_i$, Agent $i$ has a monotone sequence $T_i$ of thresholds. In order to accommodate these multiple thresholds, we only consider relations $\mathsf{R}_i \in \{ <, > , \leq, \geq \}$. 
On a play $\rho$, Agent $i$ with a multi-satisficing goal receives a \emph{payoff} (which is distinct from the cumulative reward) equal to the number of thresholds in $T_i$ that are $\mathsf{R}_i$-satisfied on $\rho$. Since the payoff structure is based on the number of thresholds met and not the actual reward assigned on a play, multi-satisficing agents always have 
a preference for a higher payoff over a lower one regardless of their relation. For example, for an agent with relation $>$ and threshold sequence $\{ 5, 10 ,15 \}$ a payoff of $0$ would be given for plays with cumulative reward $\leq 5$, representing no thresholds met. If the threshold $5$ was met but not $10$ this would be a payoff of $1$, all the way to her maximal preference of a payoff of $3$ for plays with cumulative reward greater than $15$. Even an agent with the $<$ relation would still seek to maximize the number of thresholds met; an agent with relation $<$ and thresholds $\{1,2,3\}$ would receive a payoff of $3$ on a play with cumulative reward of $0$ but a payoff of $0$ on a play with a cumulative reward of $4$.


We once again consider $W$-NEs but a modification is needed. Earlier, $W$ specified which agents had their goal satisfied. This can be seen as a list of agents that receive a payoff of $1$ as opposed to a payoff of $0$. In the multi-threshold framework, $W$ corresponds to a payoff assignment to each agent. So, if each agent had three thresholds, there would be four possible payoffs for each agent and therefore $4^k$ possible $W$'s. As an example, for an agent with threshold sequence $\{ 5, 10 ,15 \}$, if $W$ assigns a payoff of $2$, then only strategy profiles that yielded this agent a cumulative reward of more than $10$  but less than or equal to $15$ are acceptable. While more thresholds lead to more possibilities for $W$, it is often the case that a system planner has some intention towards the outcome and therefore only needs to consider $W$ that correspond to these desired outcomes. Furthermore, the different $W$'s can still be enumerated in polynomial space. 

We now construct an automaton $A_W$ that recognizes plays that assign the payoffs stipulated by $W$.  
In Section 4 we showed how to construct a B\"uchi automaton given a set of co-safety and safety automata and a specification of which automata should accept on infinite words (considering prefix satisfaction) accepted by the B\"uchi automaton. We can apply this same construction here in multiple ways by considering comparator automata that model the cumulative reward intervals corresponding to the payoffs assigned to each agent by a given $W$. For example, if we have an agent with relation $>$ and thresholds $1$ and $2$ that receives a payoff of $1$ by a given $W$, we can construct the B\"uchi automaton as in Section 4 with a co-safety comparator automaton for $> 1$ that we specify to accept along with a safety comparator automaton for $\leq 2$ that we also specify to accept -- here we are modeling payoff specification with two comparators as opposed to one. If an agent has relation $\leq$, thresholds of $0$,$1$, and $2$ and receives a payoff of $0$ then we can model this with a co-safety comparator automaton for $\leq 2$ that we specify not to accept \emph{or} a co-safety comparator for $> 2$ that we specify to accept. Since there is a choice in how to represent a given cumulative reward interval with comparators, automata may be re-used, an important direction for future work concerning implementations.

Now, we consider the deviation games constructed in Section 4.2 to analyze deviations. In order to do so, we consider the threshold representing the next highest payoff for each agent. To use our running example, if our agent with relation $>$ and thresholds $\{5,10,15\}$ was assigned a payoff of $1$ by $W$, then a play with a cumulative reward greater than $10$ represents a strictly preferable play to the current play. Therefore, we can construct a safety game (since she has a co-safety relation $>$) w.r.t the comparator that recognizes the set of plays with cumulative reward greater than $10$. This lets us reapply the construction from Section 4.2 in order to compute $\Win_1(G_j)$ -- which once again represents the set of states from which profitable deviation is possible. 

We can now construct the automaton $A'_W$ by restricting $A_W$ with respect to $\Win_1(G_j)$ as described in Section 4.3. Note that, as before, we consider games for \emph{every} agent not achieving their maximal payoff, as only agents achieving their maximal payoff do not consider deviations.

\begin{theorem}
For a given multi-agent multi-satisficing game $\mathbb{G}$, we have that $A'_W$ is nonempty iff there exists a $W$-NE strategy profile.
\end{theorem}
\begin{proof}(Sketch)
$(\rightarrow)$ Assume that $A'_W$ is nonempty -- it accepts a word $\rho \in D^{\omega}$. We construct a $W$-NE strategy profile by taking $\rho$ to be the primary trace. By construction of $A'_W$ through the comparator cross product, we have that the payoffs assigned to each agent on $\rho$ must be as stipulated by $W$. Let us now consider the case of deviations. Since no states containing $\Win_1(G_j)$ are seen along $\rho$ by construction, we remain in $\Win_0(G_j)$ states and therefore a deviation can be met with a strategy profile $\pi_0$ (the winning strategy in $G_j$ for Agent~0) that ensures the deviation fails. Therefore, by concatenating $\pi$ and the appropriate $\pi_0$ we ensure no deviations are possible.

$(\leftarrow)$ Given a $W$-NE strategy profile $\pi$, $A'_W$ accepts the sequence of decisions on the primary trace by construction.
\end{proof}

We are now able to reduce the problem of $W$-NE existence to that of non-emptiness of B\"uchi automata for multi-satisficing games, while retaining the same complexity -- $A_W$ is still exponential w.r.t the input and each deviation game is still polynomial in size, so this algorithm runs in PSPACE. 
\begin{theorem}
The problem of deciding whether a $W$-NE exists in a multi-agent multi-satisficing game is in PSPACE.
\end{theorem}

This result can also be extended, as before, to problems such as the existence of \emph{some} Nash equilibria in a multi-satisficing game by repeatedly running the algorithm for every possible $W$. Since the algorithm for deciding a single $W$ is in PSPACE, all $W$ can be checked in PSPACE as well.


\section{Relationship to $\epsilon$-Equilibria}

The notion of an $\epsilon$-equilibria modifies the standard notion of the Nash equilibria. In a Nash equilibrium strategy profile, no agent could deviate to obtain a strictly preferable payoff -  an $\epsilon$-equilibrium adds an extra condition to ``preferable". Let $\mathbb{G}$ be a discounted-sum game. A strategy profile $\sigma = \langle \sigma_0 \ldots \sigma_i \ldots \sigma_{m-1} \rangle$ is an \emph{$\epsilon$-equilibrium}, for some fixed constant $\epsilon$, in $\mathbb{G}$ if there is no agent $i \in \Omega$ with a strategy $\sigma'_i \not = \sigma_i$ such that $R_i(\rho_{\sigma}) + \epsilon \leq R_i(\rho_{ \langle \sigma_0 \ldots \sigma'_i \ldots \sigma_{m-1} \rangle})$. 

Since deviations must generate at least $\epsilon$ more cumulative reward in an $\epsilon$-equilibrium, this solution concept also addresses the problems of minuscule deviations raised in Section 3. This is one of the reasons $\epsilon$-equilibria have become an extremely popular concept in algorithmic game theory \cite{roughgarden2010algorithmic,complexityofnash}. In this general concurrent game setting solving for $\epsilon$-equilibria entails adding in extra assumptions, see \cite{guptathesis,ummelsthesis,nestochastic} for a few examples.

In this section we develop a one-way correspondence between $\epsilon$-equilibria in discounted-sum games and the $W$-NE in the multi-satisficing games of section 5. Specifically, if we are given a discounted-sum game $\mathbb{G}$ and a fixed constant $\epsilon > 0$, we show how to construct a family of multi-satisficing games of the form $\langle \mathbb{G}, \alpha \rangle$ (representing $\mathbb{G}$ plus appropriately chosen multi-satisficing goals $\alpha$, so each $\langle \mathbb{G}, \alpha \rangle$ is a multi-satisficing game) such that the existence of some $W$-NE in some $\langle \mathbb{G}, \alpha \rangle$ corresponds to an $\epsilon$-equilibria in $\mathbb{G}$. 

Here, we explore sufficient restrictions on the multi-satisficing goals $\alpha$ such that the correspondence holds. For an agent $i$ in a discounted-sum game, let $l_i$ ($g_i)$ be a lower (upper) bound on the minimal (maximal) possible cumulative reward available to the agent over all possible plays.
\begin{theorem}[$\epsilon$-Equilibria Correspondence]
Let $\mathbb{G}$ be a discounted-sum game with $m$ agents and $\epsilon > 0$ a fixed constant.  Let $\alpha = [\alpha_0, \alpha_1 \ldots \alpha_{m-1} ] $ be a multi-satisficing goal such that each $\alpha_i$ (which has smallest threshold $a_i$ and largest threshold $z_i)$  satisfies the following properties: ({\bf 1}) $\alpha_i$ has the  $\geq$ relation ({\bf 2}) $a_i \leq l_i$ ({\bf 3}) $z_i \geq g_i$ ({\bf 4}) For each pair of consecutive thresholds $t_j, t_{j+1}$ in $\alpha_i$ we have that $t_{j+1} - t_j \leq \epsilon$. 
Let $\sigma =  \langle \sigma_0 \ldots \sigma_i \ldots \sigma_{m-1} \rangle$ be a $W$-NE for some $W$ in $\langle \mathbb{G},\alpha \rangle$. Then, $\sigma$ is also an $\epsilon$-equilibrium in $\mathbb{G}$.
\end{theorem}

 \begin{proof}
Let $\sigma$ be a $W$-NE for some $W$ in a multi-satisficing game $\langle \mathbb{G}, \alpha \rangle$ with $\alpha$ satisfying the properties above. Then for each agent $i$, there are two cases. Either the agent is given its maximal payoff by $W$ in $\langle \mathbb{G}, \alpha \rangle$ or it is given less than its maximal payoff.

If the agent achieves its maximal payoff, then it satisfies the threshold $g_i$ since the highest threshold in $\alpha_i$ was at least $g_i$. Therefore, it is not possible for an agent $i$ to deviate to achieve $\epsilon$ more cumulative reward since $\epsilon  > 0 $ and $g_i$ was a strict upper bound on the total cumulative reward achievable by agent $i$ in $\mathbb{G}$. So in this scenario this agent is not capable of deviating from $\sigma$ to achieve $\epsilon$ more reward.

If the agent $i$ doesn't achieve its maximal payoff, then it satisfies some of its thresholds but not all. Due to the specification that the lowest threshold of $\alpha_i$ is less than or equal to $l_i$, we know that this lowest threshold must be satisfied on $\rho_{\sigma}$, the primary trace of $\sigma$. Therefore, we know that $\rho_{\sigma}$ assigns $i$ a cumulative reward that satisfies some threshold $t_j$ in $\alpha_i$ but not $t_{j+1}$ for some $i$. By the fourth property of $\alpha_i$, we know that $t_{j+1} - t_j \leq \epsilon$. Suppose then that agent $i$ had a strategy $\sigma'_i \not = \sigma_i$ such that $R_i(\rho_{\sigma}) + \epsilon \leq R_i(\rho_{ \langle \sigma_0 \ldots \sigma'_i \ldots \sigma_{m-1} \rangle})$ in $\mathbb{G}$. Applying this same deviation from $\sigma_i$ to $\sigma'_i$ in $\langle \mathbb{G}, \alpha \rangle $ would then correspond to deviating from a play that satisfies $t_j$ but not $t_{j+1}$ to one that additionally satisfies $t_{j+1}$. Therefore, this would be a deviation in $\langle \mathbb{G}, \alpha \rangle $ that corresponds to strictly more payoff for agent $i$, so $\sigma$ would not be a $W$-NE as assumed. We conclude then that no such agent $i$ with deviation strategy $\sigma'_i$ exists in $\mathbb{G}$.
\end{proof}

In the theorem statement, we use the $\geq$ relation since we believe it most naturally corresponds to the definition of the  $\epsilon$-equilibrium in the literature. Although it is not standard, we can define a similar minimizing condition  -- i.e. agents with the minimizing condition are not incentivized to deviate unless they receive $\epsilon$ \emph{less}  cumulative reward and correspond to agents with the $\leq$ relation in the multi-satisficing setting. This is a natural extension that lets us consider ``modified" $\epsilon$-equilibria in which agents seek to minimize total cumulative reward or a mixed game in which  some agents maximize and others minimize.

The bounds $l_i$ and $g_i$ are critical. Without the bounds in place, it may be possible for an agent to deviate to gain more than $\epsilon$ additional cumulative reward but not satisfy additional thresholds. For example, an agent $i$ may receive a cumulative reward of $r$ and be able to deviate to a play to receive $r + \epsilon$ cumulative reward, but if both of those values are lower than agent $i$'s lowest threshold then the agent has no incentive to deviate in the corresponding multi-satisficing game, breaking the correspondence between the two solution concepts. 

For this reason, we demonstrate one simple way to compute sufficient values of $l_i$ and $g_i$. For each agent $i$, let $b_i$ be the maximum possible reward available on all edges for $i$ (for biggest) and $s_i$ be the minimum possible reward over all edges (for smallest). Then,  $g_i = \sum_{j=0}^{\infty} b_i \cdot \frac{1}{\gamma^j} $ is an upper bound for the cumulative reward for agent $i$ (representing receiving the maximum reward at every time step) and similarly $l_i = \sum_{j=0}^{\infty} s_i \cdot \frac{1}{\gamma^j} $ is a lower bound. 

Theorem 7 shows a general but only one-way correspondence between the notions of $W$-NE and $\epsilon$-equilibria. A reverse correspondence, which we leave for future work,  would  be a completely characterize $\epsilon$-equilibria purely in terms of $W$-NE and would therefore be an extremely powerful result as it would allow for our efficient automata-based algorithms to find $\epsilon$-equilibria in an extremely general game setting. 

\section{Conclusion}

In this paper, we have argued for the use of satisficing goals when analyzing Nash equilibria in multi-agent concurrent discounted-sum games. There are three main advantages to consider in favor of satisficing goals.

First, satisficing goals allow for the existence of mutually beneficial equilibria that may not exist when considering optimization goals, as discussed in Section 3. Agents with optimization goals are incentivized to deviate for negligible gain, but satisficing goals address this problem. 
Second, satisficing goals allow for the use of efficient automata-based techniques. The use of these techniques allowed us to demonstrate a PSPACE upper bound. 
Furthermore, automata-based techniques allow for the use of heuristics that have been shown to be highly efficient in practice, such as BDD-based encodings \cite{BCMDH92}.
Finally, we have shown how multi-satisficing goals allow us to consider a richer payoff structure closer to that of an optimizing agent's while still addressing the negligible deviation problem and allowing for automata-based techniques. 

There are several directions available  for future work; we mention a few of particular interest here. As mentioned before, 
it is natural to consider a PSPACE-lower bound to match our upper bound presented, noting that PSPACE-hardness is  a reasonable expectation for a multi-agent problem. Furthermore, we wish to continue exploring the relationship between $W$-NEs and $\epsilon$-equilibria and consider implementations.  
\bibliographystyle{named} 
\bibliography{aaai22}

\section*{Acknowledgements}

Work supported in part by NSF grants IIS-1527668, CCF-1704883,
IIS-1830549, CNS-2016656, DoD MURI grant N00014-20-1-2787,
and an award from the Maryland Procurement Office.

\end{document}